\documentclass[pra,twocolumn,showpacs]{revtex4}

\usepackage{graphicx}
\usepackage{bm,color}

\usepackage{graphicx}
\usepackage{amsmath}
\usepackage{bm,color}
\usepackage{multirow}

\newcommand{\be}{\begin{eqnarray}}
\newcommand{\ee}{\end{eqnarray}}

\renewcommand{\theequation}{\arabic{equation}}

\begin{document}

\title{Exact projector Hamiltonian, local integrals of motion, \\
and many-body localization with symmetry-protected topological order}
\date{\today}
\author{Takahiro Orito$^{1,2}$}
\thanks{These two authors contributed equally.}
\author{Yoshihito Kuno$^{3,4}$}
\thanks{These two authors contributed equally.}
\author{Ikuo Ichinose$^1$}
\affiliation{$^1$Department of Applied Physics, Nagoya Institute of Technology, Nagoya, 466-8555, Japan}
\affiliation{$^2$Department of Physics, Graduate School of Science, Hiroshima University}
\affiliation{$^3$Department of Physics, Graduate School of Science, Kyoto University, Kyoto 606-8502, Japan}
\affiliation{$^4$Department of Physics, Graduate School of Science, Tsukuba University}

\begin{abstract}
In this work, we construct an exact projector Hamiltonian with interactions, which is given by a sum of 
mutually commuting operators called stabilizers.
The model is based on the recently studied Creutz-ladder of fermions, in which 
flat-band structure and strong localization are realized.
These stabilizers are local integrals of motion from which many-body localization (MBL) is realized.
All energy eigenstates are explicitly obtained even in the presence of local disorders.
All states are MBL states, that is, this system is a full many-body localized (FMBL) system.
We show that this system has a topological order and stable gapless edge modes exist under the open boundary condition.
By the numerical study, we investigate stability of the FMBL and topological order.
\end{abstract}


\maketitle
\section{Introduction}

Topological order and many-body localization (MBL) are most intensively studied topics in condensed
matter physics in the last decade\cite{Nandkishore,Abanin,Parameswaran_2018}.
In particular, certain studies suggest that in some systems,
all eigenstates are MBL including those at high energies \cite{Serbyn,Huse}.
This phenomenon, which is called full MBL (FMBL), is protected from disturbance due to the existence of local integrals of motion (LIOMs)\cite{Imbrie,Ros,Chandran2,Inglis,Rademaker,Monthus_2016,Goihl,Kulshreshtha,Kulshreshtha2}.
Furthermore, some FMBL systems have topologically-ordered states at all energies~\cite{Huse,Chandran,Parameswaran_2018,Bauer_2013,KOI} and under the open boundary condition (OBC),
all states exhibit degeneracy due to the presence of gapless edges modes, which exist under the OBC.
Actually, the above properties were numerically verified for some typical topological models \cite{Bahri,Decker,Kuno}.
Furthermore, the above properties, which concern to all energy eigenstates, are closely related to a concept in the modern quantum information called stabilizer \cite{Nielsen}.
Recently, study on LIOMs in topologically-ordered localized states was given, in which a notion of topological LIOMs (tLIOMs) was proposed~\cite{wahl2020}.

Stabilizers are mutually commuting local operators, and if Hamiltonian 
consists of terms including
only the stabilizers, this Hamiltonian is called projector Hamiltonian.
Obviously, such Hamiltonian commutes with all the stabilizers, and the stabilizers play a role of LIOMs if the system is FMBL \cite{Serbyn,Huse}.

The above idea might be applicable for various interesting systems, e.g.,
majorana-fermion system \cite{Kitaev,Kells,McGinley}.
However, LIOMs, stabilizers and local gapless edge modes can be constructed 
only perturbatively except for some toy models \cite{Kitaev,Fendley}.
Therefore, discovery of nontrivial models, in which LIOMs, stabilizers and 
edge modes can be constructed explicitly, is certainly welcome.
In this work, we shall present one of them based on a flat-band system.

This work is also motivated by recent theoretical studies on a fermion system in the Creutz-ladder \cite{Bermudez,Junemann,Tovmasyan,Tovmasyan2,Sun,Barbarino_2019,Zurita,KOI,Zhou,kuno2020}, which is to be realized by cold atomic gases \cite{Shin_2020} and can be constructed in photonic crystals \cite{Mukherjee}.
With specific inter and intra-chain hopping amplitudes, the system has a 
flat-band dispersion and all eigenstates are localized within a plaquette (FMBL).
Moderate interactions between fermions do not change this localization properties, 
in particular, at low fillings.
In Ref.~\cite{KOI}, we showed that the Hamiltonian of the Creutz-ladder fermion can be 
rewritten in terms of linear-combination operators of the original fermions, and 
the transformed Hamiltonian has only local terms in terms of these operators. 
And very recently, some theoretical works have reported the presence of the MBL in some flat-band systems with interactions \cite{Sharma,Danieli1,Danieli2,Danieli3}.
In this paper, we show that composites of the new operators, which appear in the Hamiltonian,
are stabilizers, and they form LIOMs.
This finding presents a new point of view for the localization properties of the Creutz-ladder model. 

The original Creutz-ladder model is understood as a projector
system rather straightforwardly.
In this work, we extend the model in a nontrivial way by adding certain specific interactions.
Even in this case, the system does not lose properties of the projector Hamiltonian.
We explicitly construct stabilizers and study the structure of the energy eigenstates. 
The Creutz ladder model represented by the stabilizers also have some symmetries, 
these symmetries seems to be related to the structure of the energy eigenstates and to the gapless edge mode under the OBC.
This study sheds light on perturbative construction of LIOMs proposed in Ref.~\cite{Serbyn}.
In addition to analytical study on the models, numerical study is quite useful to reveal physical properties of models extended by adding extra hopping terms, 
generic interactions between fermions, etc.
Stability of the topological nature and MBL can be investigated by calculating various quantities. 

This paper is organized as follows.
In Sec.~II, we explain the properties of the original Creutz-ladder model as the projector Hamiltonian, and explicitly construct gapless edge modes.
Then, we introduce interactions, with which non-trivial
entanglements between fermions are generated.
We show that stabilizers can be constructed explicitly.
By using these stabilizers, we study energy eigenstates in detail. Topological order parameter is given explicitly.
Section III gives results of numerical study on the models in Sec.~II.
Existence of the gapless edge modes is clearly shown for the 
non-interaction case, and effects of extra hopping terms, which break the topological nature of the system, are also studied.
Then, we study non-ergodic and localized properties of 
interacting models by investigating time evolution of the
system.
Section IV is devoted for discussion and conclusion.


\section{Model and LIOMs}

\subsection{Creutz-ladder model as projector system}

\begin{figure}[t]
\begin{center} 
\includegraphics[width=8cm]{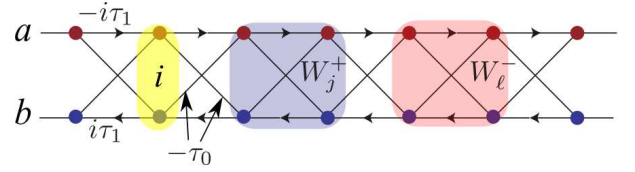} 
\end{center} 
\caption{Creutz ladder model: the yellow shade regime is a unit cell, the blue and magenta shade objects are compact localized states $W^{+}$ and $W^{-}$ defined by Eq.(6).}
\label{Fig1}
\end{figure}

In this section, we introduce a target model and explain its stabilizer structure.
Although we can directly introduce the stabilizer system, we would like to
explain its physical background.
To this end, we start with the simple Cruetz ladder model, Hamiltonian of which is given as follows,
\be
H_{\rm CL} = && \sum_j\Big[-i\tau_1 (a^\dagger_{j+1}a_j-b^\dagger_{j+1}b_j) \nonumber \\
&&\hspace{1cm} -\tau_0(a^\dagger_{j+1}b_j+b^\dagger_{j+1}a_j)+ \mbox{h.c.} \Big],
\label{HCL}
\ee
where $a_j$ and $b_j$ are fermion annihilation operator at site $j$, and $\tau_1$ and $\tau_0$ are
intra-chain and inter-chain hopping amplitudes, respectively. 
The shematic figure of the above model is shown in Fig.~\ref{Fig1}. 
In the case of $\tau_1=\tau_0$, the system has flat-band dispersions.
Furthermore, all energy eigenstates are localized in a plaquette.
To see this, it is useful to introduce the following operators,
\be 
w_{A,j}= (a_j+ib_j)/\sqrt{2}, \;\; w_{B,j}=(a_j-ib_j)/\sqrt{2},
\label{wAB}
\ee
where the above $w$-particles satisfy the fermionic commutation relations,
$\{ w_{\alpha,j}, w^\dagger_{\beta, \ell}\}=\delta_{\alpha\beta}\delta_{j\ell}$, etc.
By using $(w_{A,j},w_{B,j})$ for 
$H_{\rm flat}\equiv H_{\rm CL}|_{\tau_1=\tau_0}$,
\begin{eqnarray}
H_{\rm flat}=\sum_j2\Big[-i\tau_0w^\dagger_{A,j+1}w_{Bj}
+i\tau_0w^\dagger_{Bj}w_{A,j+1}\Big],
\label{Hflat2}
\end{eqnarray}
and therefore,
\begin{eqnarray}
&&H_{\rm flat}w^\dagger_{Aj}|0\rangle=2i\tau_0 w^\dagger_{B,j-1}|0\rangle, \nonumber \\
&&H_{\rm flat}w^\dagger_{Bj}|0\rangle=-2i\tau_0 w^\dagger_{A,j+1}|0\rangle.
\label{switch}
\end{eqnarray}
Eq.~(\ref{switch}) implies that the $w$-particle changes its species 
by the nearest-neighbor (NN)
hopping, and its location fluctuates around its original position by 
successive hoppings.
The Hamiltonian $H_{\rm flat}$ can be diagonalized by the following operators,
\be
&& W^+_j= {1 \over \sqrt{2}}(-iw_{A,j+1}+w_{B,j}),\nonumber\\
&&W^{+\dagger}_j={1\over \sqrt{2}}(iw^\dagger_{A,j+1}+w^\dagger_{B,j}), \nonumber\\
&& W^-_j= {1 \over \sqrt{2}}(-iw_{A,j+1}-w_{B,j}), \nonumber \\
&&W^{-_\dagger}_j={1\over \sqrt{2}}(iw^\dagger_{A,j+1}-w^\dagger_{B,j}), 
\label{WAB}
\ee
\be
&&W^+_j={1 \over 2}(-ia_{j+1}+b_{j+1}+a_j-ib_j), \nonumber\\
&&W^-_j={1 \over 2}(-ia_{j+1}+b_{j+1}-a_j+ib_j),
\label{Wab}
\ee
where it is easily to prove that there is one-to-one correspondence 
between $(a_j, b_j)$ and $(W^{+}_j, W^{-}_j)$ under the periodic boundary
condition (PBC), 
and $(W^{+}_j, W^{-}_j)$ satisfy the fermionic commutation relations,
\be
\{W^{\alpha \dagger}_j ,W^{\beta}_k \}=\delta_{\alpha, \beta}\delta_{jk}, 
\;\; (\alpha, \beta = \pm), \; \mbox{etc.}
\label{commW}
\ee
Under the PBC, we have
$
\sum_j[W^{+\dagger}_j W^+_j+W^{-\dagger}_jW^-_j]=\sum_j(a^\dagger_ja_j+b^\dagger_jb_j).
$

In terms of the $W$-particle, the Hamiltonian $H_{\rm flat}$ in Eq.~(\ref{Hflat2}) is diagonalized
\be
H_{\rm flat} = \sum_j [-2\tau_0W^{+\dagger}_j W^+_j+2\tau_0W^{-\dagger}_jW^-_j].
\label{HWW}
\ee
From Eqs.~(\ref{HWW}) and (\ref{commW}), it is obvious that $H_{\rm flat}$ is 
a projector Hamiltonian (stabilizer Hamiltonian), i.e.,
\be
&& [H_{\rm flat}, W^{+\dagger}_j W^+_j]=[H_{\rm flat}, W^{-\dagger}_jW^-_j]=0,  \nonumber \\
&&[W^{\alpha\dagger}_j W^\alpha_j,W^{\beta\dagger}_kW^\beta_k]=0, \;\; (\alpha, \beta = \pm).
\label{SHamiltonian}
\ee
We define stabilizers, $K^{\pm}_{0j}$, as
\be
K^{+}_{0j}=W^{+\dagger}_j W^+_j, \;\; K^{-}_{0j}=W^{-\dagger}_j W^-_j, 
\label{K0}
\ee
with properties 
\be
[K^\alpha_{0j}, K^\beta_{0k}]=0, \;\; (\alpha, \beta = \pm),
\label{K02}
\ee
and then, all energy eigenstates of the Hamiltonian $H_{\rm flat}$ are obtained from the eigenstates of 
$K^{\pm}_{0j}$ with eigenvalue $0$ or $1$.

It is straightforward to generalize the above system by introducing random
hopping amplitudes by $\tau_0\to t_j$.
Hamiltonian of the random-hopping system is given by
\be
H_{\rm rh} = \sum_j t_j[-2W^{+\dagger}_j W^+_j+2W^{-\dagger}_jW^-_j].
\label{HWWR}
\ee
Although the flat-band character is destroyed by the random hopping, the system 
$H_{\rm rh}$ in Eq.~(\ref{HWWR}) is still solvable, and all the states are localized.

Here, let us examine symmetries of the Hamiltonian $H_{\rm rh}$ [Eq.~(\ref{HWWR})]. 
In general, Hamiltonian of non-interacting fermions, $C_J=(c_J, c^\dagger_J)^t$, has the following form:
$H=\sum_{J,K} C^\dagger_J \hat{H}_{JK} C_K$, where $\hat{H}_{JK}$ is a $(2\times 2)$-matrix,
and indices $J,K$ refer to lattice site and species of fermions if there exist.
System's symmetry is defined by a unitary matrix $\Sigma_{JK}$ such as 
$\sum_{J,K}\Sigma^\dagger_{IJ}\hat{H}_{JK}\Sigma_{KN}=\pm \hat{H}_{IN}$.
System of non-interacting ferimons is classified by means of the above symmetries.
In the following, we express $\Sigma_{JK}$ by showing the transformation of $(a_j, b_j)$ under $\Sigma_{JK}$. 
There are three non-trivial symmetries.
(i) First one is closely related to topological nature
of the system, 
\be
a_j \to ib_j, \; \; b_j \to -ia_j,
\label{sym1}
\ee
which induces $W^{\pm}_j \to W^{\mp}_j$, and $H_{\rm rh} \to -H_{\rm rh}$.
As the time-reversal symmetry is broken by the complex hoppings in the Hamiltonian, 
the above symmetry (\ref{sym1}) reveals
that $H_{\rm rh}$ belongs to the AIII class of the topological insulator.
In later discussions, we shall consider additional terms.
The above symmetry plays an important role there.
(ii) Second symmetry includes the spatial inversion transformation ${\cal I}$, which induces, e.g.,
under the periodic boundary condition (PBC), $j \to -j$,
\be
a_j \to {\cal I}b_j {\cal I}=b_{-j} \;\; b_j \to {\cal I}a_j {\cal I}=a_{-j}.
\label{sym2}
\ee
The transformation (\ref{sym2}) induces $W^{\pm}_j \to \pm W^{\pm}_{-j}$ and  $H_{\rm rh} \to H_{\rm rh}$ for the case of $t_j=t_{-j}$. 
The transformation (\ref{sym2}) is nothing but the spatial $\pi$-rotation of the Creutz ladder system.
(iii) Third one is the following particle-hole type transformation,
\be
a_j \to (-)^j ib^\dagger_j, \; b_j \to (-)^j ia^\dagger_j,
\label{sym3}
\ee
which induces
$$W^+_j \to -(-)^j{W^-_j}^\dagger, \;W^-_j \to -(-)^j{W^+_j}^\dagger,$$ and  $H_{\rm rh} \to H_{\rm rh}$.
It is rather straightforward to see that the spectrum of  $H_{\rm rh}$ in Eq.~(\ref{HWWR})
is doubly degenerate by the symmetry (\ref{sym3}) [under the PBC].

Among the above symmetries, the first one plays an important role for the topological nature,
and therefore we comment on it rather in detail.
For the one-particle state $\psi_J$ with eigenenergy $\epsilon_n$, 
$\sum_J\hat{H}_{IJ}\psi_J=\epsilon_n \psi_I$, the state $\sum_J \Sigma_{IJ}\psi_J$ is also
energy eigenstate with energy $-\epsilon_n$.
This means that the quantum Hamiltonian, $H_{\rm rh}$, has the `symmetry' defined by the following
unitary operator $\Lambda$:
$ \Lambda^\dagger a_j \Lambda=ib^\dagger_j, \;   \Lambda^\dagger b_j \Lambda=-ia^\dagger_j, \;
\Lambda^\dagger a^\dagger_j \Lambda=-ib_j, \; \Lambda^\dagger b^\dagger_j \Lambda=ia_j$,
and 
\be
 \Lambda^\dagger H^\ast_{\rm rh} \Lambda = H_{\rm rh}.
\label{AIII}
\ee
As Eq.~(\ref{AIII}) shows, 
corresponding to the above single-particle symmetry, for many-body state $|n\rangle$ with
energy $E_n$, $H_{\rm rh}|n\rangle=E_n|n\rangle$, 
there exists a state with the same energy $E_n$, given by $\Lambda^\dagger |n^\ast\rangle$,
i.e., $H_{\rm rh}\Lambda^\dagger |n^\ast\rangle= E_n\Lambda^\dagger |n^\ast\rangle$.
For the state $|n\rangle$ with particle number $N$,  
the total particle number of $\Lambda^\dagger|n^\ast \rangle$ is $D_f- N$, where $D_f$ is the dimension of the Hilbert
space.
Topological invariant can be constructed from the above properties of the AIII insulator.
In Ref.~\cite{Gurarie}, it is show that topological nature of the AIII insulator is carried over to
interacting systems if the system maintains the above symmetry of the states $|n\rangle$ and $\Lambda^\dagger|n^\ast\rangle$.
Topological invariant can be constructed by Green function methods for a translationally invariant system. 

We can also add the interactions to $H_{\rm rh}$ such as
\be
H_{\rm I}= \sum c_{jk} (W^{+\dagger}_j W^+_j)(W^{+\dagger}_k W^+_k)+\cdots,
\label{Int1}
\ee
where the model is still solvable as a `classical system' with variables
$(W^{\alpha\dagger}_jW^\alpha_j)=\pm 1$.
For example, 
interactions, which respect the symmetries under (\ref{sym1})--(\ref{sym3}) (up to irrelevant constants),
give
\be
&&\sum_j\Big[W^{+\dagger}_j W^+_j+W^{-\dagger}_jW^-_j\Big]^2 \nonumber \\
&&={1 \over2} \sum_j \Big[a^\dagger_{j+1}a_{j+1}b^\dagger_jb_j
+a^\dagger_ja_jb^\dagger_{j+1}b_{j+1}\Big]   \nonumber \\
&&+\mbox{(scattering terms)}
\ee
where $\mbox{(scattering terms)}= \sum_j(a^\dagger_{j+1}a^\dagger_jb_{j+1}b_j)$, etc. which induce
scattering of $a$ and $b$-particles.

From the above construction of the model, 
$$H_{\rm tot}=H_{\rm rh}+H_{\rm I},$$
there exists a $Z_2$ topological order parameter such as 
\be
Z_2=(-1)^{\sum_j(W^{+\dagger}_j W^+_j+W^{-\dagger}_jW^-_j)}.
\label{Z2}
\ee
Here, since $[Z_2,H_{\rm tot}]=0$ holds, both $H_{\rm tot}$ and $Z_{2}$ can be diagonalized simultaneously. 
It should be remarked that ${Z_{2}}$ operator is invariant under the transformations  (\ref{sym1})--(\ref{sym3}).
As we show shortly, gapless edge modes emerge under the open boundary condition (OBC), and 
these modes generate degenerate states with different $Z_2$-quantum number, $Z_2=\pm 1$.
We can also define a string-order parameter,
$O_{\rm st}(i,k)$, which is a hallmark of the topological state.
It is given by 
$O_{\rm st}(i,k)=(-1)^{\sum^k_{j=i}(W^{+\dagger}_j W^+_j+W^{-\dagger}_jW^-_j)}$, and 
again $O_{\rm st}(i,k)$ is invariant under the transformations  (\ref{sym1})--(\ref{sym3}).
In a specific state and disorder
realization, the string-order parameter takes random values
$O_{\rm st}(i,k)=\pm 1$.

In particular, the above $Z_2$ order parameter plays an important role for 
the system with the OBC. 
In this case, 
$Z_2 \to (-1)^{\sum^{L}_{j=0}(a^\dagger_{j}a_j+b^\dagger_{j}b_j)}$.
It is interesting to see if gapless edge modes appear in the system
$H_{\rm tot}$
under the OBC as a hall mark of the topological state.
We expect that the edge modes are fermionic, and then, the degenerate states belong to
the different $Z_2$ sectors.
To see the edge modes, we consider the ladder of length $L$ with 
$j=0, 1, \cdots, L$, and define
\be
H_{\rm op} =&& \sum_{j=0}^{L-1} t_j[-2W^{+\dagger}_j W^+_j+2W^{-\dagger}_jW^-_j] \nonumber \\
&& +\sum_{j,k=0}^{L-1}c_{jk} (W^{+\dagger}_j W^+_j)(W^{+\dagger}_k W^+_k)+\cdots.
\label{Hop}
\ee
From the fact that $W^\alpha_j$ is an operator defined on the plaquette with $j$-th and $(j+1)$-th rungs,
$W^{\pm}_{-1} \to {1 \over 2}(-ia_0+b_0)$ as $a_{-1}=b_{-1}=0$ due to the present OBC. 
Then, an edge operator is expected to be 
$G_0\equiv (a_0+ib_0)/\sqrt{2}=\omega_{A,0}$. 
Here, since $\{G^\dagger_0, G_0\}=1, (G_0)^2=0$, $G_0$ is fermionic. 
Furthermore, even for the interacting model of $H_{\rm op}$, this operator
$G_{0}$ satisfies four criteria of the gapless (fermionic) edge mode
\cite{Fendley,Kitaev}: (I) The $G_{0}$ commutes with 
$H_{\rm op}$, 
$[H_{\rm op},G_0]=0$, (II) The $G_0$ anti-commutes with $Z_2$, 
(III) $\{G_0^{\dagger}, G_0\}=1$, 
(IV) $G_{0}$ acts on the edge portion of the system.
Especially, the criterion (II) guarantees that $G_{0}$ is a mapping operator
between even and odd sectors characterized by the $Z_{2}$
operator.
From the above,
it is obvious that $G_0$ represents the gapless edge mode 
dictated by the topological order.
Similarly for the $j=L$ edge, $W^{\pm}_{L} \to {1 \over 2}(a_{L}-ib_{L})$ as $a_{L+1}=b_{L+1}=0$, and then
$G_{L}=(a_{L}-ib_{L})/\sqrt{2}=\omega_{B,L}$ forms the gapless edge mode.
By the analytical calculation with $H_{\rm op}$ in Eq.~(\ref{Hop}), 
we can explicitly verify  the above conclusion.

However, 
careful examination is required to see if the above operators of the gapless modes under the OBC
really produce physically meaningful states.
Operation of $G_0, \; G^\dagger_0$ to the energy eigenstates may vanish them.
In fact, the practical calculation shows,
\be
G_0W^{+\dagger}_0|0\rangle=G_0W^{-\dagger}_0|0\rangle=G_0W^{+\dagger}_0W^{-\dagger}_0|0\rangle=0, 
\label{GW}
\ee
where $|0\rangle$ is the empty state.
On the other hand, application of the creation operator $G^\dagger_0$ produces
the state $G^\dagger_0|0\rangle={1 \over \sqrt{2}} (a^\dagger_0-ib^\dagger_0)|0\rangle$ and
\be
G^\dagger_0W^{+\dagger}_0|0\rangle=
{1 \over \sqrt{2}}[ (a^\dagger_0-ib^\dagger_0)(ia^\dagger_1+b^\dagger_1)
+2ia^\dagger_0b^\dagger_0]|0\rangle,
\ee
and similarly for $G^\dagger_0W^{-\dagger}_0|0\rangle$ and 
$G^\dagger_0W^{+\dagger}_0W^{-\dagger}_0|0\rangle$.
All the states reside on the $j=0$ plaquette are doubly degenerate,
and these degenerate states have different $Z_2$-values with each other.

In the later study by means of the numerical methods in Sec.~III,
we consider additional hopping terms [Eq.~(\ref{HIR})].
Among them, 
first one [$H_{\rm IR1}$] respects the symmetry under Eq.~(\ref{sym1}) [and also Eq.~(\ref{sym2})],
whereas second [$H_{\rm IR2}$]  does not.
In both cases, $G_0$ is not the genuine gapless edge mode.
However in the first case, we can construct a gapless mode 
whose leading contribution comes from $G_0$.
Therefore, {\em the present system can have a symmetry protected topological state}.

From the view point of MBL, $K^{\pm}_{0j}$ is nothing but LIOMs, the existence of which indicates
that all energy eigenstates are localized.
In the present system, inclusion of arbitrary hoppings of $a$ and 
$b$-particles break the above
LIOM structure, but it is plausible to expect that the system maintains its
localization properties as in the ordinary one-dimensional systems.
Our recent study on the Creutz-ladder revealed that the inclusion of random on-site potentials
induces a crossover from the flat-band localization to the Anderson-type localization
as a result of the breaking of the above LIOM picture. 

It is interesting to extend the above discussion on the projector Hamiltonian
with LIOMs
to interacting systems that have non-trivial LIOMs.
Stability of MBL coming from LIOMs, then, can be investigated.
This will be studied in the following subsection.

\subsection{Interacting models}

In Sec.~IIA, we considered the genuine flat-band Creutz-ladder model of fermions.
There, the stabilizers $\{K^\alpha_{0,j}\}$ have only 
single-plaquette entanglements, and therefore, each
energy eigenstate is a simple product state.
From this aspect, $\{K^\alpha_{0,j}\}$ have properties
of $\ell$-bit operators 
({\em local-bit}
operators)~\cite{Serbyn,Huse,Parameswaran_2018}.

In this subsection, we shall introduce a projector
Hamiltonian with non-trivial interactions and
entanglements.
The interaction, which we consider, is given as follows,
\be
H_{\rm N}=\sum_j gN_{j-1}[W^{+\dagger}_jW^-_j+W^{-\dagger}_jW^+_j],
\label{HN}
\ee
where $N_j\equiv [W^{+\dagger}_jW^+_j+W^{-\dagger}_jW^-_j]=K^+_{0j}+K^-_{0j}$, and $g$ is
an arbitrary parameter.
Total Hamiltonian of the target model is $H_{\cal T}=H_{\rm rh}+H_{\rm N}$. 
The reason to consider the interaction $H_{\rm  N}$ in Eq.~(\ref{HN}) will be explained shortly.

We shall show that the system $H_{\cal T}$ is a projector Hamiltonian with topological order.
We first seek stabilizers by extending $K^\alpha_{0j}$ of the Creutz-ladder
model.
To this end, we calculate the commutator between $K^\alpha_{0j}$ and 
$H_{\rm N}$ to obtain
\be
[K^+_{0j}, H_{\rm N}]=gN_{j-1}\{W^{+\dagger}_jW^-_j-W^{-\dagger}_jW^+_j\}.
\label{com1}
\ee
On the other hand, we introduce the following operators $O_j$,
\be
O_j=N_{j-1} \{W^{+\dagger}_jW^-_j+W^{-\dagger}_jW^+_j\},
\label{Oj}
\ee
and practical calculation shows
\be
[O_j, H_{\rm rh}]=4t_jN_{j-1}\{W^{+\dagger}_jW^-_j-W^{-\dagger}_jW^+_j\}.
\label{OjH}
\ee
From Eqs~(\ref{com1}) and (\ref{OjH}), we can prove
\be
[K^+_{0j}-{g\over 4t_j}O_j, H_{\cal T}]=0.
\label{Kstab1}
\ee
Similar calculation to the above shows
\be
[K^-_{0j}+{g\over 4t_j}O_j, H_{\cal T}]=0.
\label{Kstab2}
\ee
Then we define operators $K^\alpha_j \; (\alpha=\pm)$ as, 
\be
&& K^+_j\equiv K^+_{0j}-{g\over 4t_j}O_j, \nonumber \\
&& K^-_j\equiv K^-_{0j}+{g\over 4t_j}O_j.
\label{Kstab3}
\ee
Practical calculation shows that $K^\alpha_j \; (\alpha=\pm)$ commute with each other,
$[K^\alpha_j, K^\beta_k]=0$, and therefore, they are stabilizers.
Finally, we can prove 
\be
H_{\cal T}=\sum_j2t_j(-K^+_j+K^-_j).
\label{HTKK}
\ee
Here we should stress that the above results are satisfied exactly for arbitrary $t_j$'s and $g$, 
as a result of the specific choice of the interaction $H_{\rm N}$ in Eq.~(\ref{HN})~\cite{FN1}.
More complicated systems with Hamiltonian composed of products of the stabilizers 
are also solvable.
But they may exhibit different behaviors from $H_{\cal T}$ in Eq.~(\ref{HTKK})
concerning to dephasing and temporal developing of entanglement
entropy, etc., if non-local (long-range) terms of the stabilizers exist in the Hamiltonian.


\subsection{Eigenstates and topological order}

In this section, we study the energy eigenstate of the Hamiltonian $H_{\cal T}$ in Eq.~(\ref{HTKK}).
This study is useful for later discussion on the topological order, in
particular, existence of gapless edge modes.
Energy eigenstates are constructed from eigenstates of the stabilizers as a product state.

To investigate eigenstates of the stabilizer somewhat in detail, 
let us study quantum states residing on site (plaquette) $j$ as a concrete example.
From Eqs.~(\ref{Oj}) and (\ref{Kstab3}), the states at site $j$ have non-trivial entanglement
with the state at site $(j-1)$.
We first introduce notations for the states;
\be
&&W^+_j|0\rangle_+=0, W^{+\dagger}_j|0\rangle_+=|1\rangle_+, \nonumber \\
&&W^-_j|0\rangle_-=0, W^{-\dagger}_j|0\rangle_-=|1\rangle_-.
\ee
For the case of $\langle N_{j-1}\rangle=0$, 
$K^\alpha_j=K^\alpha_{0j}=W^{\alpha \dagger}_jW^\alpha_j$,
and then, the eigenstates of $K^+_j$ are $|0\rangle_+$ and $|1\rangle_+$ with
eigenvalue $0$ and $1$, respectively.
Similarly for $K^-_j$.
In particular, the $\langle N_j\rangle=1$ sector such as $N_j|\psi \rangle=|\psi\rangle$ is spanned by the states 
$\{|1\rangle_+|0\rangle_-, |0\rangle_+|1\rangle_-\}$.

On the other hand, the cases of $\langle N_{j-1}\rangle=1$ and $2$ are non-trivial.
Let us set the stabilizers as 
\be
&&K^+_j=K^+_{0j}-\lambda \{W^{+\dagger}_jW^-_j+W^{-\dagger}_jW^+_j\},  \nonumber \\
&&K^-_j=K^-_{0j}+\lambda \{W^{+\dagger}_jW^-_j+W^{-\dagger}_jW^+_j\},
\label{KN1}
\ee
where $\lambda=g/(4t_j) \; (g/(2t_j))$ for $\langle N_{j-1}\rangle=1 \; (2)$.
Eigenstates of the above stabilizers are obtained as follows for the $\langle N_j\rangle =1$ sector,
\be
&&\psi_1\propto x_1 |1\rangle_+|0\rangle_--\lambda |0\rangle_+|1\rangle_-, \nonumber \\
&&\psi_2 \propto \lambda  |1\rangle_+|0\rangle_-+x_1|0\rangle_+|1\rangle_-,
\label{eigen}
\ee
where 
\be 
&&K^+_j \psi_1=x_1\psi_1, \; \; K^+_j\psi_2=x_2\psi_2,  \nonumber \\
&&K^-_j \psi_1=x_2\psi_1, \; \; K^-_j\psi_2=x_1\psi_2, 
\label{eigenK} 
\ee
with $x_1={1 \over 2}(1+\sqrt{1+4\lambda^2})$ and  $x_2={1 \over 2}(1-\sqrt{1+4\lambda^2})$.
The above result obviously indicates that unitary rotation in the two-dimensional vector space 
$\{|1\rangle_+|0\rangle_-, |0\rangle_+|1\rangle_-\} \to \{\psi_1, \psi_2 \}$ is induced by the interaction.
This unitary rotation is given by operator such as 
\be
U(\theta)=e^{\theta (W^{+\dagger}W^--W^{-\dagger}W^+)},
\label{Utheta1}
\ee
with a suitable angle $\theta$.
Here, $[W^{+\dagger}W^--W^{-\dagger}W^+)]^2=-1$ for the present sector [$N_j=1$],
and therefore
\be
U(\theta)=\cos \theta +\sin \theta \cdot (W^{+\dagger}W^--W^{-\dagger}W^+).
\label{Utheta}
\ee
From Eqs.~(\ref{eigen}) and (\ref{Utheta}), value of $\theta$ is determined for $\psi_{1(2)}$,
i.e., $\tan \theta=\lambda/x_1$.
In Ref.~\cite{Serbyn}, methods of constructing LIOMs was discussed.
There, unitary rotation of subsystem plays a central role.
The above discussion shows that the present system is one of examples of
the LIOMs construction given in Ref.~\cite{Serbyn}.

Finally for the sectors with $\langle N_j\rangle=2$ and $\langle N_j\rangle=0$, the expression of the stabilizers $K^{\pm}_j$
in Eq.~(\ref{KN1}) shows that the $\lambda$-dependent terms are irrelevant,
and the eigenstates of the stabilizers are simply
$\{|1\rangle_+|1\rangle_-, |0\rangle_+|0\rangle_-\}$.

Let us turn to the system with the OBC, $j=0, \cdots, L$.
In this case, Hamiltonian of the OBC is derived from $H_{\cal T}$, and 
for the $j=0$ boundary, the corresponding term is given by
$2t_0(-K^+_{0j=0}+K^-_{0j=0})$ as $\langle N_{-1}\rangle =0$.
Then, the discussion on the edge modes given in the previous section is directly applied to
the $j=0$ boundary and we find that all states in the $j=0$ plaquette are doubly degenerate.
On the other hand for the $j=L-1$ plaquette, the full stabilizers appear in the Hamiltonian
$H_{\cal T}$.
By fixing the $W$-particle number at $j=L-2$, the above discussion on the eigenstates 
can be applied to the $j=L-1$ plaquette.
There, we show that the existence of the interaction term induces the rotation 
in the $\{W^+, W^-\}$ operator plane for the $\langle N_{L-1}\rangle =1$ case. 
Then by using the rotated $W$'s, similar argument to the case of $g=0$ can be applied to
the interacting case.
However from Eq.~(\ref{WAB}), a rotation of $\{W^+, W^-\}$ means simply a change of relative weight 
of $w_{A,j+1}$ and $w_{B,j}$.
This indicates that the edge mode $G_L=(a_L-ib_L)/\sqrt{2}$ in the $g=0$ case
also works in the interacting case.
Practical calculation shows, 
$$
\{G^\dagger_L, W^+_{L-1} \}=\{ G^\dagger_L, W^-_{L-1} \}=0,
$$
then, degenerate states are obtained by applying the operator $G^\dagger_L$ 
to the states at $j=L$.

If we introduce interactions expressed in terms of $K^+_j$ and $K^-_j$ such as
$\sum c_{jk}K^+_jK^+_k + \cdots$, there appear correlations between states residing on
different sites.
Even in this case, the above gapless edge modes with the OBC survive.
On the other hand for general interactions, which are {\em not written in terms of}
$K^+_j$ and $K^-_j$, energy eigenstates may not be constructed by simply
applying the unitary rotation $U(\theta)$ in Eq.~(\ref{Utheta1}).

The topological order for $H_{\cal T}$ is also defined as the previous case for $H_{\rm tot}$, and the suitable order parameter characterizing the above
topological states is a `generalization' of Eq.~(\ref{Z2}) and ${O}_{\rm st}(i,k)$, 
\be
&& \tilde{Z}_2=(-1)^{\sum_{j,\alpha}K^\alpha_j}, \nonumber \\
&& \tilde{O}_{\rm st}(i,k)=(-1)^{\sum^k_{j=i}(K^+_j+K^-_j)},
\label{Z22}
\ee
where we should note $[K^\alpha_j]^2 \neq K^\alpha_j$ in general, but from Eq.~(\ref{Kstab3}),
$K^+_j+K^-_j=K^+_{0j}+K^-_{0j}=W^{+\dagger}_jW^+_j+W^{-\dagger}_jW^-_j$. 
$\tilde{Z}_2$ operator also commutes with $H_{\cal T}$,  $[\tilde{Z}_2,H_{\cal T}]=0$.
As there exist zero-energy edge modes,
each eigenstate are doubly degenerate for each $Z_{2}$-sector.
Furthermore, even if we add  interactions expressed in terms of $K^+_j$ and $K^-_j$ to $H_{\cal T}$,  
$ \tilde{O}_{\rm st}(i,k)$ operator acts as $ \tilde{Z}_2$-topological order parameter, 
and $Z_{2}$-topological order exists.

Finally, let us comment on the symmetries of the interacting model.
It is not so difficult to show that the system, $H_{\cal T}$ [Eq.~(\ref{HTKK})],
has  properties similar to the non-intercation one which is explanied in Sec.~II.A.
In fact as shown in the above, $(-K^+_j+K^-_j) \psi_{j1}=(-x_1+x_2)\psi_{j1}$, whereas
$(-K^+_j+K^-_j) \psi_{j2}=(x_1-x_2)\psi_{j2}$.
The empty state $|0\rangle$ degenerates with the total full-state, 
$|{\rm F}\rangle =\prod_{\mbox{\tiny all} \; j}\psi_{j1}\psi_{j2}$, and the lowest-energy state (for the case $t_j>0$) is the half-filled state
$|{\rm LE}\rangle=\prod_{\mbox{\tiny all} \; j}\psi_{j1}|0\rangle_{j-}$.
(We expect that notations are self-evident.)
Therefore, the interacting systems, $H_{\cal T}$, carry over the properties of the AIII class insulator,
and the symmetry is $Z_2\times Z^f_2$, where $Z^f_2$ is the fermion-number parity.
More precisely, there exists a counterpart $|n^\ast\rangle$ for each energy eigenstate $|n\rangle$,
but their exact degeneracy is broken by the interaction.
However, the chiral properties of $\psi_J$ and $\sum_{J}\Sigma_{IJ}\psi_J$ in the single-particle system 
are carried over to the interacting system, i.e., for each $|n\rangle$ with $E_n$, there exists $|\tilde{n}\rangle$
with $-E_n$.
We expect that this $Z_2$ symmetry plays a similar role as the single-particle $Z_2$ symmetry, and
as a result, there exist the gapless edge modes and the nonlocal string order as we saw in the above.
According to the group super-cohomology classification of fermionic 
symmetry protected topological (SPT) phases~\cite{classSPT1,classSPT2}, 
the present interacting models have the $Z_2$-SPT phases in the spatial one-dimension.
Detailed numerical studies are welcome to verify this observation.



\section{Numerical studies}

In Sec.~II, we introduced models and studied their physical properties concerning MBL and
topological order.
In this section, we shall study the target systems in detail by employing numerical 
exact diagonalization~\cite{Bukov2}.

\subsection{Edge modes in random Creutz-ladder models}
We first study the edge modes existing in the random Creutz-ladder model with the OBC.
We calculate the energy eigenvalues of the model Eq.~(\ref{HWWR}) by fixing random hoppings $\{t_j\}$ and verify that there are two gapless edge modes.
It is interesting to see (in)stability of the edge modes in the
existence of other type of hopping terms.
To study this problem, we consider the following two kinds of 
inter-chain hopping,
\be
H_{\rm IR1}&=&v_1\sum_j(a^\dagger_jb_j+b^\dagger_ja_j)\nonumber \\
              &=&v_1i\sum_j(w^\dagger_{A,j}w_{B,j}-w^\dagger_{B,j}w_{A,j}), \nonumber \\
H_{\rm IR2}&=&v_2i\sum_j(a^\dagger_jb_j-b^\dagger_ja_j)  
\nonumber \\
              &=&v_2\sum_j(w^\dagger_{A,j}w_{A,j}-w^\dagger_{B,j}
              w_{B,j}),
\label{HIR}
\ee
where $v_1$ and $v_2$ are real parameters.
We use the $w_{\rm A,B}$-particle representation as 
the gapless edge modes in the OBC are nothing but
$G_0=w_{A,0}, \ G_L=w_{B,L}$, and therefore, the topological nature of the system becomes clear 
in the description by the $w_{A, B}$-particle.
We call the system given by the Hamiltonian $H_{\rm rh}+H_{\rm IR1, 2}$ extended Creutz-ladder model,
where $H_{\rm rh}$ in Eq.~(\ref{HWWR}) is expressed as 
$
H_{\rm rh}=\sum_j2t_j\Big[-iw^\dagger_{A,j+1}w_{Bj}
+iw^\dagger_{Bj}w_{A,j+1}\Big].
$
In the practical calculation, we sometimes employ 
spin-$1/2$ operators as $w^{(\dagger)}_{A(B),j}\to S^{-(+)}_{A(B)j}$,
$n^{A(B)}_j-1/2\to S^z_{A(B),j}$ related by a Jordan-Wigner (JW) transformation. 
As numerical bases, it is sufficient to use the product states of spin-$z$ component base for each sites.
Since in the $w$-particle Hamiltonian under the JW transformation, 
the total spin-$z$ component in the system is conserved.
Then, in practical calculation, we fix the total spin-$z$ component 
to reduce the numerical cost.

We display the energy eigenvalues for the case of the uniform and random 
hoppings in Appendix A.
For the uniform-hopping case, addition of
$H_{\rm IR1}$ destroys the flat-band structure, but gapless edge modes survive.
On the other hand by adding $H_{\rm IR2}$, the flat-band structure remains 
but gapless edge modes split into two in-gap edge modes with energy $\pm \Delta \propto \pm v_2$~\cite{kuno2020}.
For the random-hopping case, similar results are obtained, 
in particular, for the gapless edge modes.
In fact in the presence of $H_{\rm IR1}$, the gapless edge 
modes are obtained perturbatively.
It is straightforward to verify that 
$\tilde{G}_0=\omega_{A,0}+{v_i\over t_0}\omega_{A,1}
+{v^2_1 \over t_0t_1}\omega_{A,2}+\cdots$ commutes with
the total Hamiltonian, $H_{\rm rh}+H_{\rm IR1}$.
In the existence of $H_{\rm IR1}$, 
the ``chiral symmetry'' in Eq.~(\ref{sym1}) is not broken
and zero-energy edge modes exist. 
Therefore in the present system, the chiral symmetry protects
gapless edge modes.

\begin{figure}[t]
\begin{center} 
\includegraphics[width=8cm]{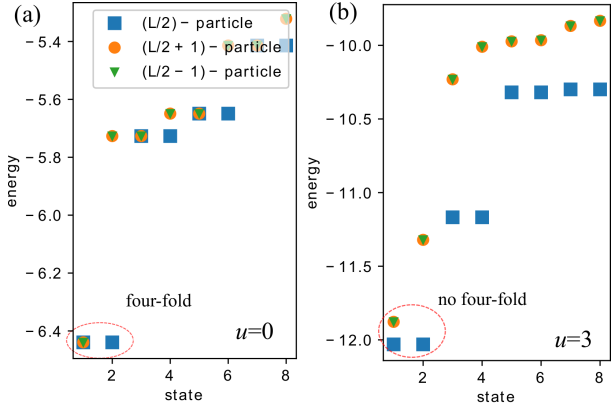} 
\end{center} 
\caption{Single-shot disordered many-body spectra
for different fillings, the data show the many-body energy
from the lowest to the eighth states. 
Here, $2t_j=1+\delta_j$ with the uniform distribution
random values
$\delta_j\in[-0.5,0.5]$. (a) $u=0$.
There exist four-fold degenerate states with different
particle numbers indicating the existence of zero-energy
edge modes.
(b) $u=3$. The four-fold degeneracy is broken.
The system size is $L=14$.}
\label{Fig2}
\end{figure}

We examine the density profile of the $w_{\rm A,B}$-particles in
the many-body wave function to verify that the gapless modes
are really localized in the edges. 
In terms of the many-body wave function, the signal of the presence of some localized
zero-energy edge modes may be a ``filling anomaly''~\cite{bibo2019,you2020}.
For example, for our model with the OBC and in the vicinity of the half filling,
if left and right localized edge modes are present in the system, 
$L/2-1$, $L/2$, and $L/2+1$ particle states are energetically degenerate. 
This indicates emergence of four-fold degenerate states.
Actually we demonstrate it numerically for 
the non-interacting case with $v_1=v_2=0$.
In Fig.\ref{Fig2} (a), the results are for the ground-state and
low-lying excitation energies for different particle number 
near the half filling. 
From the results of the three different fillings in 
Fig.\ref{Fig2} (a), four-fold degenerate states certainly emerge.

Furthermore, we demonstrate how the edge modes 
change by adding interactions. 
As shown in the previous section, even in some specific 
interacting case, the local stabilizer was found as 
in Eq.~(\ref{HTKK}). 
Therefore, the degeneracy by means of edge modes obviously
exists in that system.
Here, we focus not on the specific form of Eq.~(\ref{HTKK}) 
but on another type of interactions.
In the $w$-particle representation, we consider the following interactions,
\be
H_{\rm S}=
\sum_{j}\frac{u}{4}\biggl[(n^A_{j}-{1\over 2})
(n^B_{j}-{1\over 2}) 
+(n^B_{j}-{1\over 2})(n^{A}_{j+1}-{1\over 2})\biggl], \nonumber \\
\label{Hint_w}
\ee
where $n^{A(B)}_j=w^{\dagger}_{A(B),j}w_{A(B),j}$.
It should be remarked here that the interactions in Eq.(\ref{Hint_w}) as well as
the hopping amplitudes
in Eqs.~(\ref{HIR}) cannot be expressed in terms of
the stabilizers $K^+_{0j}$ and $K^-_{0j}$, and therefore, 
the system described by 
the total Hamiltonian loses the projector nature and the edge 
modes tend to unstable for sufficiently large $u$. 
Figure \ref{Fig2} (b) displays the typical result for it. 
The four-fold degenerate states existing for $u=0$ split into 
three states with different particle filling for $u=3.0$. 
Therefore, a ``phase transition" to a non-topological state
takes place there,
We numerically estimate the critical coupling, $u_c$, as
$u_c\sim 2$. We note that similar dynamics have been reported in a diamond flat band system with interactions, 
where fate of the localization dynamics was studied \cite{Marco}.

\begin{figure}[t]
\begin{center} 
\includegraphics[width=8cm]{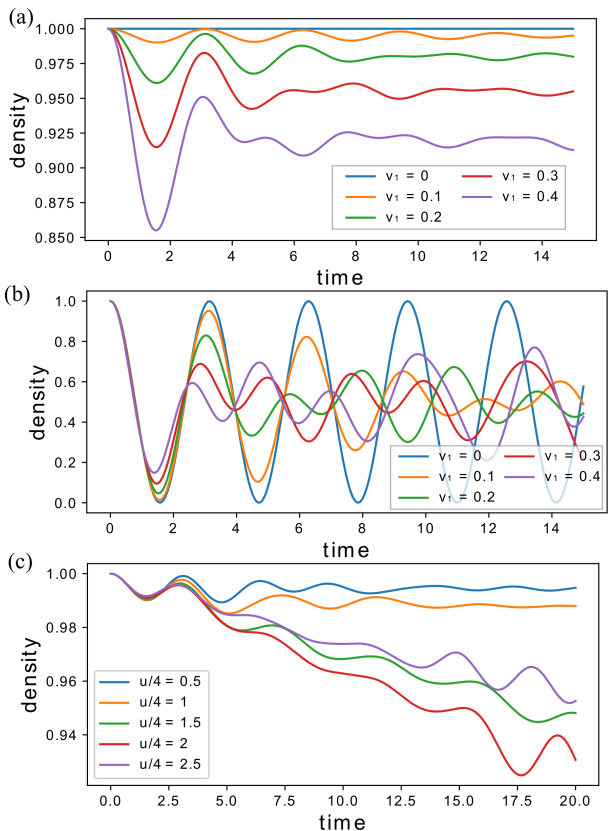} 
\end{center} 
\caption{Short-time exact dynamics for the 
$H_{\rm flat}+H_{\rm IR1}$ system. 
(a) $P_0(t)$ in Eq.~(\ref{P0t}). 
(b) Dynamics of the $w$-particle density at the center of
the system. System size $L=8$.
(c) Short-time exact dynamics of $P_{0}(t)$ in the 
presence of interaction of $H_{S}$ and weak $H_{\rm IR1}$.
Here, $v_1=0.1$, $L=8$. 
For all data, the unit of time is $\hbar/(2\tau_0)$, and 
unit of energy $2\tau_0$.}
\label{Fig3}
\end{figure}

Next, we shall examine the stability of edge modes by studying time evolution 
of the states that include the edge modes.
In particular, we focus on effects of the inter-chain hopping, 
$H_{\rm IR1}$ in Eq~(\ref{HIR}), on the stability. 
That is, we focus on $H_{\rm flat}+H_{\rm IR1}$.

As the initial state for the time evolution, we consider the following states,
\be
|\Psi_0\rangle=G^\dagger_0|\Phi_q\rangle, \;\; |\Psi_L\rangle=G^\dagger_L|\Phi_q\rangle,
\label{Psi0}
\ee
with
\be
|\Phi_q\rangle = \prod^{2qL}_{i=1} A_q|0\rangle, 
\label{Phi0}
\ee
where 
$
A_q=w^{\dagger}_{A,(2q)^{-1}(i-1)+1},
$
and $q$ is given by $q=1/2$ for the density-wave like bulk state,
which we consider in the following.
By numerical methods, we obtain the wave function at $t$, $\Psi(t)$, and 
calculate
the following quantity to observe the (in)stability of the edge modes,
\be
&&P_0(t)=\langle \Psi_0(t)|G^\dagger_0G_0|\Psi_0(t)\rangle \label{P0t}
\label{edgeob}
\ee
If $P_0(t)$ 
decreases from unity, the edge modes lose their 
coherence [coherence of $|a\rangle$ and $|b\rangle$].
On the other hand, if it keeps a value close to unity for large $t$, 
the edge modes and topological nature of the system are stable 
against the decoherence.

Numerical results are shown in Fig.~\ref{Fig3} for various values
of $v_1$. There, we calculate $P_{0}(t)$ and also the particle
density at typical sites in the bulk. 
We consider the case of the uniform hopping for the flat-band localization
as we are first interested in the stability of edge modes, 
behavior of the bulk states, and effects of the interaction $H_{\rm S}$ in the genuine system. 
Study on the random systems is a future work.
Figure~\ref{Fig3}(a) shows that as $v_1$ increases, the value of 
$P_0(t)$ is getting small for the initial stage of the time evolution, but it has a fairly large value
for $t \to$ large for all $v_1$.
This result strongly indicates the existence of stable zero modes in the vicinity of the edges.
On the other hand, Fig.~\ref{Fig3}(b) shows
that the $w$-particle density at the central regime of the system
oscillates around $0.5$ for any value of $v_1$. 
Hence, the dynamics of the localized edge states is clearly
different from that of the bulk dynamics.

In addition to the effects of $H_{\rm IR1}$, 
we investigate the effects of the interaction of Eq.~(\ref{Hint_w})
with a weak but finite $v_1=0.1$, i.e., the model is 
$H_{\rm flat}+H_{\rm IR1}+H_{S}$. 
The results are displayed in Fig.~\ref{Fig3}(c). 
The stability of the localized edge states is gradually lost as
increasing the interaction $u$.
On the other hand, for the $u/4=2.5$ case, the decay is 
somewhat weakened.
Possible reason for this behavior is that the entanglement spreading 
is faster compared to the other cases, and the spreading saturates 
in a finite system. 
Anyway, detailed analysis is interesting and this is a future work.


\subsection{(Non)ergodicity of interacting model}

In this subsection, we shall study MBL for the system $H_{\cal T}$ 
[Eq.~(\ref{HTKK})] including the NN interactions.
Even in the presence of the interaction of $H_{\rm N}$ 
[Eq.~(\ref{HN})], 
the total Hamiltonian can be described by the local stabilizer form. 
Accordingly, if we consider the dynamics for a certain quantum state based on this system, the information of the quantum state can be conserved. 
Here, as a typical example, we investigate time evolution where the initial states is given by Eq.~(\ref{Phi0}). 
To this end, as explained in the previous subsection, the
$w$-particle representation is employed and for 
$H_{\cal T}$, we use
\be
&&N_j=(w^\dagger_{A,j+1}w_{A,j+1}+w^\dagger_{B,j}w_{B,j}), \nonumber \\
&&W^{+\dagger}_jW^-_j+W^{-\dagger}_jW^+_j
=(w^\dagger_{A,j+1}w_{A,j+1}-w^\dagger_{B,j}w_{B,j}).
\nonumber
\ee
To study the dynamics, we calculate the expectation value of
the particle density in the central regime of the system
as a function of time, which is
set to unity in the initial state.

\begin{figure}[t]
\begin{center} 
\includegraphics[width=8cm]{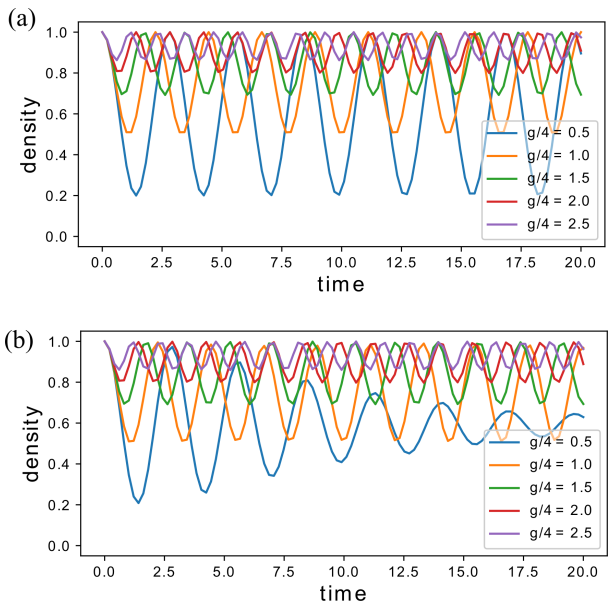} 
\end{center} 
\caption{Short-time exact dynamics of the particle density
at the center of the system.
(a) Dynamics for the system $H_{\cal T}$.
Data show perfect oscillating behavior of all values of $g$.
Period of the oscillation is longer for smaller $g$.
(b) Dynamics for the system
$H_{\cal T}+H_{\rm IR1}$. Here, $v_1=0.1$. 
Particle density exhibits stable oscillation for larger value of $g$, whereas it decreases
to $1/2$ for small $g$. 
For all data, $L=8$. 
The hopping amplitude in $H_{\cal T}$ is set to be uniform: $t_j=\tau_0$. 
The unit of time is
$\hbar/(2\tau_0)$, and unit of energy $2\tau_0$.}
\label{Fig4}
\end{figure}

Figure~\ref{Fig4} is the result of the time evolution of the particle density at the center of the system.
For Fig.~\ref{Fig4}(a), the clear and perfect oscillating behavior is
observed for any value of $g$, which indicates the
particle-density revivals. 
This means that the
information of the density distribution at the initial state
is preserved, and therefore, the ergodicity is broken. 
Furthermore, we add $H_{\rm IR1}$ and investigate the effect
of $H_{\rm N}$.
Here, we set $v_1=0.1$ and employ the same initial state and 
calculate the same observable in Fig.~\ref{Fig4} (a). 
The result of the dynamics is shown in Fig.~\ref{Fig4} (b).
Interestingly enough, even for the presence of  $H_{\rm IR1}$ term, 
the revival oscillation gets recovered as
increasing the strength of $H_{\rm N}$. 
This shows that as increasing the interaction, the nature 
as stabilizer of the projector Hamiltonian $H_{\cal T}$
gets dominant in the dynamics of the system.
Strong correlations between NN plaquettes generated by
$H_{\rm N}$ enhances the non-ergodicity.
Very recently, interesting observation concerning to
oscillating behavior in MBL states, a quantum time crystal,
was given by using notion of extensive dynamical 
symmetries \cite{Medenjak}.
The above numerical results suggest that they exist
in the present system, $H_{\cal T}$ [Eq.~(\ref{HTKK})].
This is an interesting future problem.

\section{Conclusion}

In this work, we studied the Creutz-ladder model of fermions
and its extensions from the view point of the projector system.
It is known that the Creutz-ladder model has a flat-band 
dispersion and exhibits the strong localization properties 
for specific choice of the intra-chain and inter-chain couplings.
We revealed that the model with the above hopping amplitudes
has a projector Hamiltonian.
The stabilizers are explicitly obtained.
We showed that 
they play a role of LIOMs, which dictate localization of all
energy eigenstates.
Furthermore, we showed that the systems have topological nature
and there exist gapless edge modes under the OBC.
Numerical study indicates that the above single-particle 
properties are preserved even in the existence of interactions
up to moderate coupling constants.

Next, we generalized the Creutz-ladder model with preserving
its projector nature.
The new model has strong correlations between particles in
NN plaquettes.
The stabilizers are explicitly obtained and they work on particles
in NN plaquettes.
This generalized model retains the topological nature, and 
there exist gapless edge modes.
Typical energy eigenstates are analytically obtained, and they give
an important insight on construction of LIOMs in generic systems.

Finally, by numerical methods, we investigated time-evolution
of the states with and without edge modes in order to examine
the (non)ergodicity of the system. 
The inter-chain hopping amplitude, which breaks the chiral
symmetry, plays an important role for the time evolution.

Our findings in the present work are quite interesting and
useful for the investigation of the full localization and 
topological order from the view point of the projector Hamiltonian.
The present models are feasible by recent experiments on cold
atomic gases, and we hope that our findings will be verified by 
experiments.
In particular, the time evolution of the systems is very interesting
and important for the quantum information physics.

\section*{Acknowledgments}
This work is supported by the Grant-in-Aid for JSPS Fellows (No.17J00486).

\appendix
\setcounter{figure}{0}
\def\theequation{\thesection.\arabic{equation}}
\def\thefigure{\thesection.\arabic{figure}}
\setcounter{section}{0} 
\section{Single-particle spectrum of various extended models of Creutz-ladder fermion system}
\begin{figure}[h]
\begin{center} 
\includegraphics[width=8.4cm]{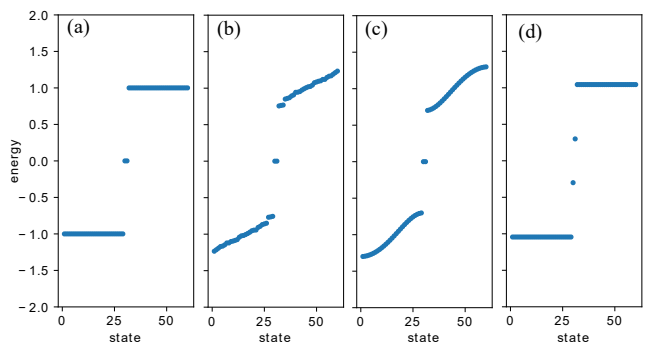} 
\end{center} 
\caption{Energy spectra in the $w$-particle representation
under the OBC.
(a) Uniform coupling $2t_j=1$. 
Flat-band dispersion and 
zero-energy modes exist in the band center.
(b) Single-shot disordered spectra for random coupling: $2t_j=1+\delta_j$ with the uniformly-distributed random
values $\delta_j\in[-0.5,0.5]$. 
(c) $v_1=0.3$.
Flat bands disappear, but zero-energy modes exist in 
the band center.
(d) $v_2=0.3$.
Flat bands exist, but zero-energy modes split into 
inter-band states with finite energy.
}
\label{FigA1}
\end{figure}

As we explain in the main text, we study a single-particle spectrum for various extended
models of the Creutz-ladder fermions.
Through this investigation, we can observe relevant terms in the Hamiltonian for preserving/ breaking 
topological character of the Creutz-ladder model with the specific hopping amplitude, $\tau_0=\tau_1$.

We consider the OBC, and results are shown in Fig.~\ref{FigA1}.
Fig.~\ref{FigA1} (a) is the energy spectrum for the non-random uniform hopping amplitude with $\tau_0=\tau_1$.
There exist two flat bands and two energy-zero modes between the flat bands.
Figure~\ref{FigA1} (b) shows the result of the random hopping amplitude, and there still exist two gapless modes.
The above two cases are analytically studied in the main text and the numerical results are what we expected.
Figure~\ref{FigA1} (c) and (d) are results for the models with uniform coupling $\tau_1=\tau_0$ and including inter-chain hopping such as 
$(v_1=0.3, v_2=0)$ and $(v_2=0.3, v_1=0)$, respectively.
These results are interesting.
In the case (c), there still exist two gapless modes in the band center.
On the other hand in the case (d), two flat bands are preserved, whereas the inter-band modes 
split into two gapful modes.

From the above observation, we conclude that the chiral symmetry 
given by the transformation of Eq.~(\ref{sym1}) is essential ingredient for the existence of the gapless modes.
In the main text, we show that these gapless modes are confined near the boundaries of the system.




\begin{thebibliography}{99}
\bibitem{FN1}
It is interesting to consider symmetry of the projector Hamiltonian with non-trivial interactions.
In fact, the Hamiltonian $H_{\rm N}$ in (\ref{HN}) transforms 
as $H_{\rm N} \to H_{\rm N}$ under Eq.~(\ref{sym1}).
We can also consider another projector Hamiltonian such as
$H'_{\rm N}=i\sum_j gN_{j-1}[W^{+\dagger}_jW^-_j-W^{-\dagger}_jW^+_j]$, 
which transform as $H'_{\rm N} \to -H'_{\rm N}$ under Eq.~(\ref{sym1}).
Topological properties, e.g., existence of gapless modes and the string order parameter are the same
for both systems, $H_{\rm rh}+H_{\rm N} [H'_{\rm N}]$.
\end{thebibliography}
\end{document}